\documentclass[showpacs,twocolumn,aps,amssymb,preprintnumbers,superscriptaddress,prl,floatfix]{revtex4-1}

\usepackage{epsfig,dcolumn}

\usepackage{graphicx,bm}

\usepackage{bm,color}

\usepackage{hyperref}

\newcommand{\be}{\begin{equation}}

\newcommand{\ee}{\end{equation}}

\newcommand{\ba}{\begin{eqnarray*}}

\newcommand{\ea}{\end{eqnarray*}}

\begin{document}

\title{Shape Coexistence in  $^{78}$Ni as the Portal to the Fifth Island of Inversion}

\author{F. Nowacki}

\affiliation{Universit\'e de Strasbourg, IPHC, 23 rue du Loess 67037 Strasbourg, France}
\affiliation{CNRS, UMR7178, 67037 Strasbourg, France}

\author{A. Poves}

\affiliation{Departamento de F\'isica Te\'orica e IFT-UAM/CSIC, 
Universidad Aut\'onoma de Madrid,  E-28049 Madrid, Spain\\
and Institute for Advanced Study,  Universit\'e de Strasbourg}

\author{E. Caurier}

\affiliation{Universit\'e de Strasbourg, IPHC, 23 rue du Loess 67037 Strasbourg, France}
\affiliation{CNRS, UMR7178, 67037 Strasbourg, France}

\author{B. Bounthong}

\affiliation{Universit\'e de Strasbourg, IPHC, 23 rue du Loess 67037 Strasbourg, France}
\affiliation{CNRS, UMR7178, 67037 Strasbourg, France}

\begin{abstract}

Large Scale Shell Model calculations (SM-CI) predict that the region of deformation which 
comprises the heaviest Chromium and Iron isotopes at and beyond N=40 will merge 
with a new one at N=50 in an astonishing parallel to the N=20 and N=28 case in the
Neon and Magnesium isotopes. We propose a  valence space including the
full $pf$-shell for the protons and the full $sdg$ shell for the neutrons, which represents a come-back of the
the harmonic oscillator shells  in the very neutron rich regime. 
The onset of deformation is
understood in the framework of the algebraic  SU3-like structures linked to quadrupole dominance.
Our calculations preserve the
doubly magic nature of the ground state of $^{78}$Ni, which, however,  exhibits  a well deformed prolate band at low 
excitation energy,  providing a striking example of shape coexistence far from stability. This new IoI adds to the four well
documented ones at N=8, 20, 28 and 40.

\end{abstract}


\date{\today}

\maketitle

\textit{Introduction.}
         The limits of the  nuclear stability  and  the origin  of the chemical elements in the universe
         are  burning issues in nuclear physics.
        These fundamental questions pose a major challenge to our
        theoretical understanding of nuclei, in particular towards the neutron drip line, given the important
        role played by the structure of neutron-rich
        nuclei  in numerous  astrophysical processes \cite{zn82mass}. 
        A  peculiar manifestation of the nuclear dynamics, absent in
        most interacting fermionic systems, is the phenomenon of shape coexistence, i.e, the
        appearance of quantum states of very different shapes very close in energy, typically
        much smaller than what could be anticipated in a mean field description \cite{hw}. This is a 
        major consequence of the dominant role of the correlations in the nucleus.
        The experimental exploration of  regions of very neutron rich  nuclei has recently 
        unveiled the presence of a rich variety of shape coexistence scenarios in this sector
        of the nuclear chart. They are  particularly frequent at the edge of the so called \mbox{Islands of Inversion, (IoI's)}
        which appear close to  semi-magic or even doubly magic nuclei.  The IoI's are formed when a group of nuclei,
        expected to be spherical in their ground states, become deformed, precisely because 
        the strong nuclear quadrupole-quadrupole interaction produces a shape transition in which highly
        correlated many particles many holes configurations dubbed "intruders" turn out to be
        more bound than the spherical
        ones. These intruder deformed bands often appear at low excitation energy in the magic nuclei
        close to the IoI's, hence, shape coexistence  acts  in fact as a portal to the IoI's.
        An intense experimental program 
        \cite{Sorlin2002,Hannawald,Rother,Ljungvall,Gade10,dijon,recchia,suchyta,baugher,crawford}
        has recently explored the details of the shape coexistence in $^{68}$Ni and the
        existence of a new IoI surrounding $^{64}$Cr.    
         Large scale shell model calculations predicted first,  and later explained in detail this behaviour \cite{lnps,mcsm}, 
         with particular emphasis in the strong similarity of the mechanisms driving the onset of deformation and the collapse of the shell closures
         at N=20 and N=40. Another recent experimental and theoretical finding is the merging of the IoI's
         at N=20 and N=28 \cite{Doornenbal,2028}. 
        In view of these precedents, 
         we wondered if  a new IoI might exist at N=50 and if   an equivalent phenomenon of IoI's merging
         occurs for N=40 and N=50, and these questions prompted this study,
         whose results support fully both conjectures. In the meantime, there have been new experimental measures on the chromium  and iron isotopes
         up to N=42 and N=46 respectively, which seem to support it as well \cite{santamaria},  and interesting theoretical an experimental explorations
          of the physics close
         to  $^{78}$Ni \cite{fred_kam,ge80int,zn79-is}, which have helped us a lot in the present investigation. 
          The trends in the evolution of the spherical mean field which 
         favour the appearance of  the IoI's  have been discussed in ref. \cite{SorlinINPC}.     
        
         \textit{The valence space.} Why should intruder
          configurations produce shape coexistence in  $^{78}$Ni and deformed ground states
          in  $^{76}$Fe and    $^{74}$Cr?  Because of     
          the interplay of the spherical mean field and the quadrupole-quadrupole
          correlations. Therefore, we must ensure that the chosen 
          valence space can incorporate the latter properly.  
            Due to the similarity between the N=20-28 
             and the  N=40-50 regions, we chose to define an
             interaction covering the {\it pf-sdg} valence space which parallels
            what we did successfully for the  {\it sd-pf}  case \cite{iokin,2028}.
          It is well known that quadrupole collectivity is maximised when the 
          orbits close to the Fermi level permit the realisation of two variants of Elliott's SU(3)
            \cite{neqz,pseudo1,pseudo2}; Pseudo-SU(3) which occurs when all the orbits of a major oscillator shell except the one with the largest {\bf j}
           are quasi-degenerated, and Quasi-SU(3) when the orbit  with the largest {\bf j} and its  $\Delta$j=2 $\Delta$l=2  partners are 
           quasi-degenerated. 
           For the protons, the natural space beyond Z=20 and up to Z=32-34 is indeed the full $pf$ 
          shell,  which will bring-in quadrupole collectivity of Quasi or Pseudo-SU(3) type 
          depending on the number of protons
           and the value of the Z=28 gap. 
          In the neutron side, as we add neutrons in excess of N=40, the role of the neutron 
          $pf$-shell orbits is slowly transferred to the 
          companions of the 0g$_{\frac{9}{2}}$ orbit in the $sdg$ major oscillator shell. 
          Unfortunately, size prevents us to following the transition 
          from N=40 to N=50 keeping both sets of neutron orbits in our valence space, but we expect a smooth transit from one space to another. 
          The border line between the space of ref.~\cite{santamaria} and the present 
          one can be loosely located at N=46.  Just at the transition point it could be advisable to include the 1p$_{\frac{1}{2}}$ neutron orbit 
          as well.  In the N=40 valence space, the
          neutron collectivity is provided by the Pseudo-SU(3) coherence of the holes below N=40 and the  Quasi-SU(3) coherence of the
          particles above it, and the mean field regulator is the N=40 gap. Close to N=50 the build-up of collectivity is different. 
          Neutrons above N=50
          can develop Pseudo-SU(3) coherence, whereas 
           neutron  holes  in   the  0g$_{\frac{9}{2}}$  neutron orbit  help with their single 
           shell prolate quadrupole  moment.
          The mean field regulator is  now the N=50 gap. In brief, the valence space  proposed in this article comprises
          the p=3 harmonic oscillator (HO,  E(p)=$\hbar \omega (p+1/2)$) 
          shell ($pf$) for the
          protons and the  p=4 HO shell ($sdg$) for the neutrons.
          Thus, compared to that of ref. \cite{lnps} the new valence space corresponds to closing the $pf$ shell and including the missing $sdg$ shell
          orbits for the neutrons,  using an inert core of  $^{60}$Ca.

\begin{figure}
\begin{center}
\includegraphics[width=\columnwidth]{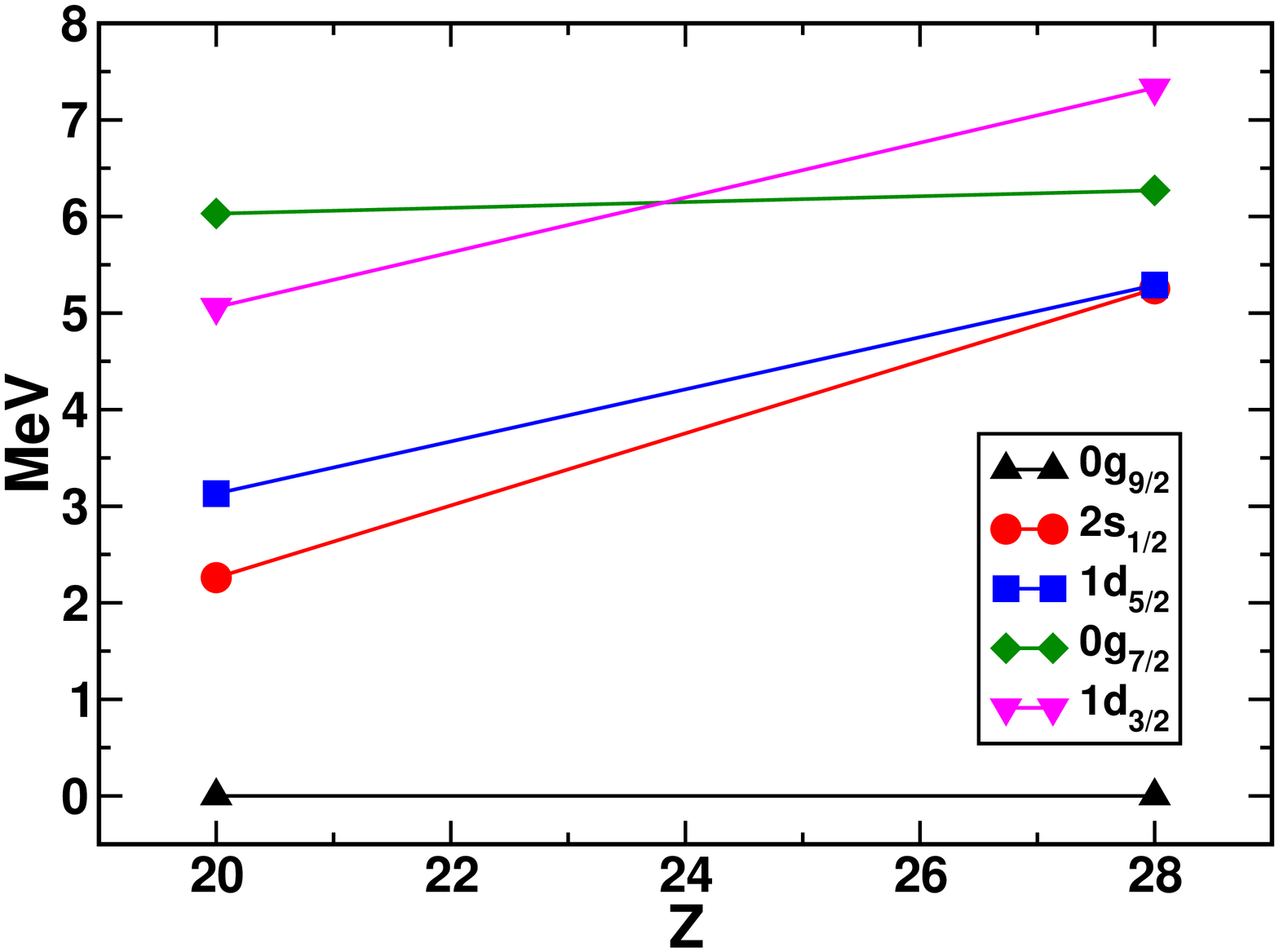}
\includegraphics[width=\columnwidth]{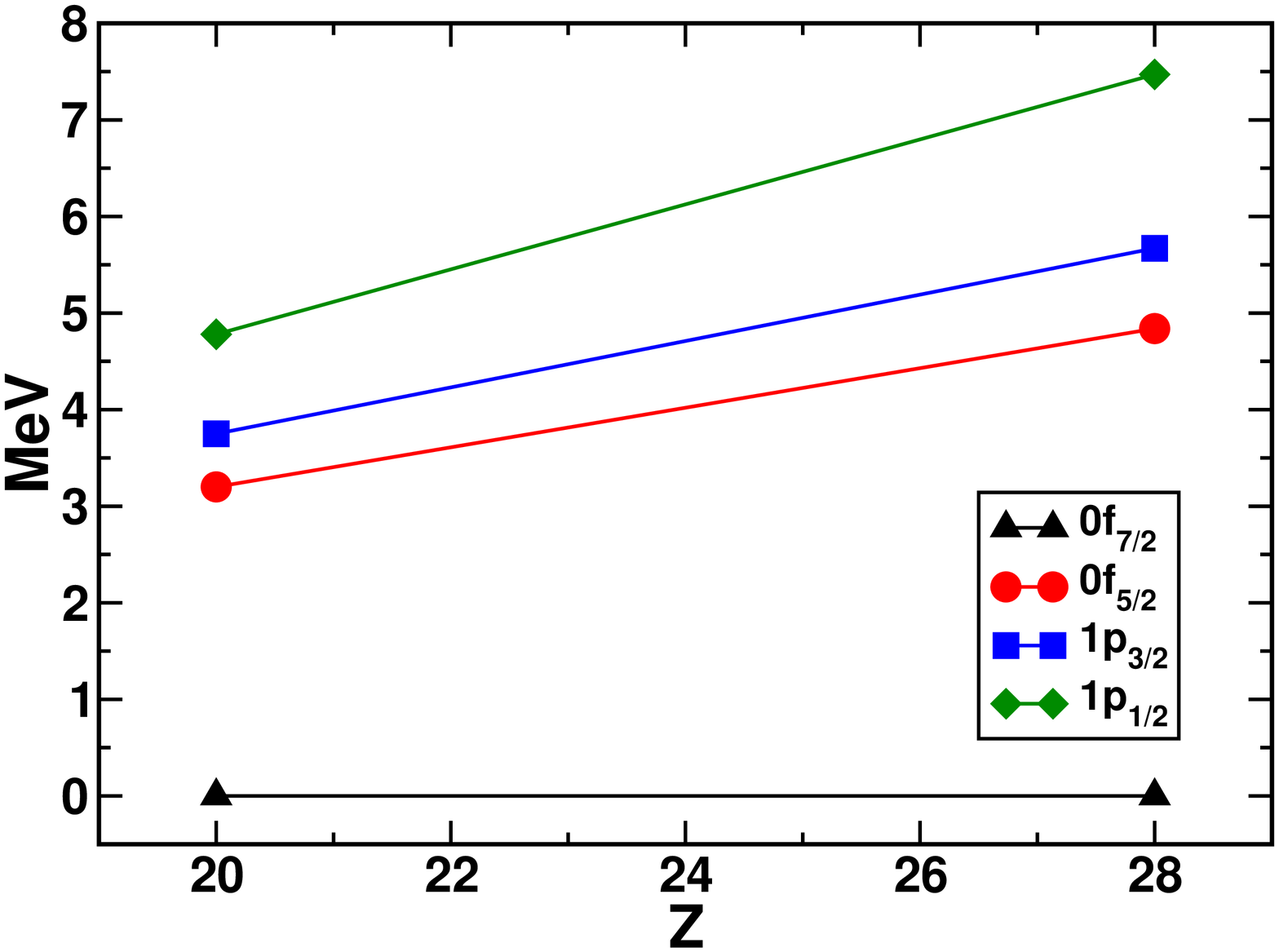}
\caption{(color online)  Effective Single Neutron Energies at N=50 (upper panel) and
 Effective Single Proton Energies at N=50 (lower panel). \label{espe}}
\end{center}
\end{figure}

         \textit{The effective interaction.} Our starting effective interaction matrix elements  are based on the
          free nucleon-nucleon interaction of reference \cite{cdbonn}, regularised and renormalised with the  G-matrix techniques of 
           reference \cite{morten}. We modify their monopole part in order to reproduce the experimental evolution 
          of the regulating gaps in the model space.
          Following the previous study of $^{78}$Ni in \cite{fred_kam} 
          we have left unchanged most of the  monopoles of the
          {\sc lnps} interaction and, 
          to fix the remaining ones, we have adjusted the neutron ESPE's to reproduce the experimental 
         binding energies of the nickel isotopes ($S_n$ for A=69 and $S_{2n}$ for A=68-72),
         the binding energies of the zirconium isotopes for A=90-98 and 
         the neutron gaps in the  N=50 isotones.
          The size of the Z=28 proton gap is controlled  by the B(E2)'s
          from the first excited 2$^+$ states to the 0$^+$ ground states of the zinc and germanium 
          isotopes around N=50. We adopt for $^{79}$Cu the single particle spectrum of 
           the {\sc jun45} interaction \cite{JUN45}. Finally, 
           the $^{79}$Ni single particle spectrum and the cross-shell
           $pf$-$sdg$ proton-neutron monopoles are constrained to
           reproduce the N=49 and N=51 available data. In particular
           the relative location of the 9/2$^+$, 5/2$^+$ and  1/2$^+$ states
          in the N=49 isotones and their evolution from
           Z=30~\cite{zn79-is} to  Z=32-40~\cite{nndc}.
          The final values of  the Effective Single Particle  Energies
           are gathered  in Figure \ref{espe}.
           We name the resulting
           interaction {\sc  pfsdg-u}. The dimensions of the matrices to be diagonalized are at the edge of present computer
           capabilities for solving the
           problem exactly, in several cases reaching 2$\times$10$^{10}$ Slater determinants.

\begin{figure}[t]
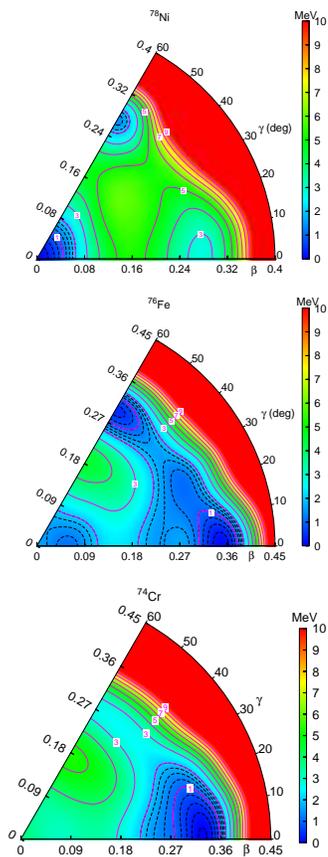

\begin{center}
\includegraphics[width=0.5\columnwidth]{ni78.eps} \\ 
\includegraphics[width=0.5\columnwidth]{fe76.eps}\\ 
\includegraphics[width=0.5\columnwidth]{cr74.eps}
\caption{(color online) Projected  Energy Surfaces for $^{78}$Ni, $^{76}$Fe, and  $^{74}$Cr with the interaction   {\sc  pfsdg-u}.
\label{N=50}}
\end{center}
\end{figure}

\textit{Projected Energy Surfaces.} 
To unveil  the presence of  intrinsic structures underlying  our SM-CI results in the laboratory frame,
we have performed Constrained  Hartree Fock (CHF) calculations in the $pf-sdg$ valence space using the 
 {\sc  pfsdg-u} interaction. 
We solve the  Hartree-Fock equations constrained by  the quadrupole deformation parameters $\beta$ and $\gamma$
--which break rotational invariance and thus produce deformed shapes -- to obtain  the energy as a function of  these parameters.
The results are displayed as  contour plots in the 
(mass) $\beta$-$\gamma$ plane (aka Projected Energy Surfaces PES) in Fig.~\ref{N=50}. 
It is seen that at the mean field level, $^{78}$Ni
is an spherical doubly-magic nucleus in its ground state with local deformed oblate and prolate minima.  
$^{76}$Fe  has instead three  minima,  spherical, prolate 
 and oblate, the latter two connected by the $\gamma$ degree of freedom. 
  In the case of $^{74}$Cr, the landscape is fully dominated by the
 prolate  solution. These deformed structures have typically $\beta$$\sim$0.3.
 
 \textit{Spectroscopic results.} Let's move now to the predictions of the full fledged diagonalizations  using the interaction {\sc  pfsdg-u}, starting with
          the results at fixed number of neutron excitations across the N=50 closure. For this calculations we do not impose any truncation in the
          proton space.
           The  structure of the 2p-2h and
           4p-4h bands are very similar for all the isotopes (except for  $^{70}$Ca) 
          and correspond to well deformed rotors with a nearly perfect J(J+1) spacing and B(E2)'s  consistent with deformation
           parameters very close to the ones obtained in the SU3 limit  (we use standard effective charges q$_{\pi}$=1.31 
           and q$_{\nu}$=0.46).  
           For the 2p-2h yrast band of  $^{74}$Cr we have E(2$^+$)=0.27~MeV and 
           B(E2)(2$^+_1$$\rightarrow$ 0$^+_2$)= 360~e$^2$fm$^4$, whereas for the 4p-4h one we get
           E(2$^+$)=0.17~MeV and 
           B(E2)(2$^+_1$$\rightarrow$ 0$^+_2$)= 555~e$^2$fm$^4$. 
            We have estimated the correlation energies of the 2p-2h and 4p-4h
           neutron configurations, diagonalizing a properly normalized quadrupole interaction in the {\it sdg} space for the neutrons and the Quasi-$pf$
           doublet for the protons.  The results are displayed in Table~\ref{tab:qqcorr}. It is seen that both for the 2p-2h and 4p-4h cases the
           largest correlation energies correspond to $^{74}$Cr and $^{76}$Fe  followed by those of  $^{78}$Ni and $^{72}$Ti. Notice that removing 
           protons from $^{78}$Ni, the intruder configurations will benefit from the gain in correlation energy  {\bf and}  from the reduction of the N=50
           neutron gap, therefore we may expect an abrupt shape change producing an IoI.
           
\begin{table}
\caption{Quadrupole correlation energies of  the neutron intruder configurations, relative to the N=50 closure
 (in MeV). \label{tab:qqcorr} }   
\begin{tabular*}{\linewidth}{@{\extracolsep{\fill}}|c|ccccc|}
\hline 
  & $^{78}$Ni &  $^{76}$Fe &  $^{74}$Cr  &  $^{72}$Ti  &  $^{70}$Ca  \\
  \hline 
 2p-2h & 5.3 & 6.5 & 7.0   & 5.3  & 2.2  \\
  4p-4h & 9.3 & 10.9 & 11.3   & 9.1  & 4.8 \\          
 \hline    
\end{tabular*}
\end{table}

 \begin{figure}
\begin{center}
\includegraphics[width=\columnwidth]{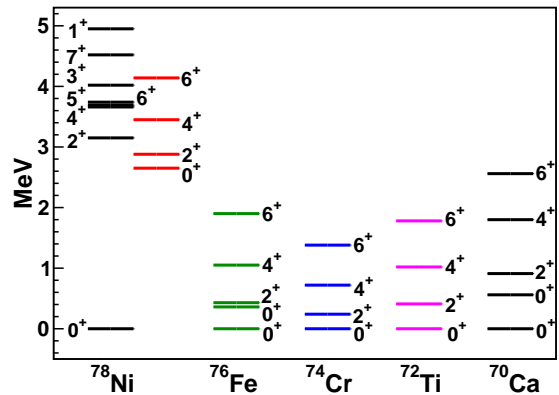}
\caption{(color online).  Theoretical spectra of  the N=50 isotones
with the {\sc pfsdg-u}  interaction.  In red the  deformed intruder band of $^{78}$Ni. \label{n50spec}}
\end{center}
\end{figure}
\begin{table}
\caption{Some E2 properties of the N=50 isotones. 
Energies in MeV,
B(E2)'s in e$^2$fm$^4$, Q's in efm$^2$. \label{tab:be2} }  
\begin{tabular*}{\linewidth}{@{\extracolsep{\fill}}|c|ccccccccc|}
\hline 
  &  \multicolumn{3}{c}{$\Delta E$}  &\multicolumn{3}{c}{B(E2)$\downarrow$} & \multicolumn{3}{c}{Q$_s$}\\
  \hline
  &  ${2^+}$&  ${4^+}$  &${6^+}$  &2$^+$ & 4$^+$  & ${6^+}$  & 2$^+$ & 4$^+$ & ${6^+}$  \\
  \hline   
 $^{78}$Ni & 2.88 & 3.45 & 4.14  & 32  & 783 & 1021 & -39  & -65 & -75  \\
  $^{76}$Fe & 0.43 & 1.05 & 1.90 & 314  & 707 & 780  & -45  & -57 & -63  \\
   $^{74}$Cr & 0.24 & 0.72 & 1.38 & 630  & 911 &1004  & -51 & -66 & -74  \\
    $^{72}$Ti & 0.41 &1.02 & 1.78& 321  &506  & 580&  -34 & -45 & -53  \\
     $^{70}$Ca & 0.91& 1.80 & 2.56 & 119 & 194  & 5 & -3  & +8 & +8   \\
           \hline      
\end{tabular*}
\end{table}      
\begin{table}
\caption{Average number of p-h excitations and occupancies of the neutron and proton orbits above N=50 and Z=28 for several intruder states \label{tab:occup} }   
\begin{tabular*}{\linewidth}{@{\extracolsep{\fill}}|c|cc|cccc|ccc|}
\hline 
     & $n_{ph}^{\nu}$  & $n_{ph}^{\pi}$ &  $d_{5/2}^{\nu}$    & $s_{1/2}^{\nu}$    & $g_{7/2}^{\nu}$    & $d_{3/2}^{\nu}$    & $p_{3/2}^{\pi}$    & $f_{5/2}^{\pi}$    & $p_{1/2}^{\pi}$   \\ \hline
 $^{78}$Ni 0$^+_2$ & 2.7 & 2.3  &  1.1  & 0.8   & 0.4 & 0.4 & 0.9 & 1.0 & 0.4 \\
  $^{76}$Fe 2$^+_1$ &  3.0 & 1.4 & 1.2  & 0.8   & 0.6 & 0.4 & 0.8 & 0.4 & 0.2 \\
   $^{74}$Cr 0$^+_1$  & 4.9 & 1.6  &   1.8  & 1.1   & 1.2 & 0.8 & 1.1 & 0.3 & 0.2 \\
   $^{72}$Ti  0$^+_1$ & 4.8 & 0.9 & 2.2  & 0.7   & 0.6 & 1.3 & 0.7 & 0.1 & 0.1 \\
      $^{70}$Ca 0$^+_1$ & 3.5 & 0.0 &  1.9  & 0.3   & 0.2 & 1.1 & 0.0 & 0.0 & 0.0 \\
     \hline 
 \end{tabular*}
\end{table}

         For the full diagonalizations we use a  truncation  scheme in terms of the sum of the number of  neutron excitations across  N=50 and 
          proton excitations across Z=28 (t).  We perform full space
          calculations for Ca , Ti  and Cr  and   we are limited to t=8  for Ni and Fe, but  the calculations seem to be converged. 
          For   $^{78}$Ni (see Fig.~\ref{n50spec}) we predict a doubly magic  ground state at 65\%, 
          with a first 2$^+$
          excited state at 2.88~MeV,  which belongs to  the  (prolate) deformed band based in the
            intruder 0$^+$  which appears at  an excitation energy of  2.65~MeV  and a second  2$^+$ of 1p-1h nature  at 
           3.15~MeV,  connected to the ground state with B(E2)=110~e$^2$fm$^4$. We have plotted as well 
           the yrast 4$^+$ which belongs to the deformed band, its  6$^+$ member, and several  states of particle-hole nature.
           \mbox{The B(E2)(2$^+_1$$\rightarrow$ 0$^+_2$) 
           goes up to 516~e$^2$fm$^4$.}  The location of the intruder band depends of the competition of the
           monopole losses whose linear part is given by the neutron ESPE's and the correlation gains (see Table~\ref{tab:qqcorr}).
            In $^{78}$Ni the balance favours  the
           closed shell, with the intruder 2p-2h (neutron) band below 3~MeV.  Removing two 
          protons, in  $^{76}$Fe,  the N=50 gap is reduced and the correlation energy increased. This produces an abrupt lowering of the
            intruder configurations whose  band-heads become nearly degenerated with the 0p-0h N=50 closure.  Hence, the
            ground state of $^{76}$Fe turns out to be a very complicated mixture of np-nh configurations, including 21\% of 0p-0h
            and 33\% of neutron 2p-2h. 
           The yrast  2$^+$ appears at  0.43~MeV and it is rather of 2p-2h plus 4p-4h nature. This mismatch produces a certain quenching of the
           B(E2) relative to the spectroscopic quadrupole moment of the 2$^+$ as seen in Table~\ref{tab:be2}.
           Most interestingly,  the first excited state is another 0$^+$  at 0.36~MeV which is also of very mixed nature, although now the
            0p-0h components amount to 33\%.  The distortion of the spectrum is due to the mixing of the spherical and the deformed  0$^+$'s.
            Thus,  the doublet of 0$^+$ states in $^{76}$Fe signs  the rapid  transition
            from the doubly magic ground state of  $^{78}$Ni to the fully rotational case of  $^{74}$Cr, where  the collective behaviour
           is well  established  and  the neutron 4p-4h intruder becomes dominant in the yrast band,  
           with a 2$^+$ at  0.27~MeV
           and E(4$^+$)/E(2$^+$)=~3 (see Figure~\ref{n50spec}).  Collectivity persists to a lesser extent in  $^{72}$Ti, whose  2$^+$ is 
           at  0.41~MeV.  There is no experimental 
           information for these nuclei 
           yet.    Table~\ref{tab:be2}  shows 
           the calculated B(E2) values and spectroscopic quadrupole moments, 
           which correspond, in the well deformed case of   $^{74}$Cr, to
           $\beta_{mass}$$\sim$0.32 and
           $\beta_{charge}$$\sim$0.35 in very nice agreement with the results of the CHF-PES. In Table~\ref{tab:occup} we display the
           occupation numbers of the neutron and proton orbits above the N=50 Z=28 doubly magic closure. It is seen that in the neutron side 
           they evolve from 
           2.7 neutrons excited in  $^{78}$Ni, to a maximum of 4.9 neutrons in  $^{74}$Cr, and down to 3.3 neutrons in  $^{70}$Ca. Importantly, we verify
            that in all the cases, all the excited orbits have non-negligible occupations, as expected in a Pseudo-SU3 regime, which, however, is
            only fully dominant in $^{74}$Cr. In the proton sector, the p$_{3/2}$ orbit is preferentially populated, as it should happen in the Quasi-SU3 limit,
            except in   $^{78}$Ni where the proton collectivity is rather of Pseudo-SU(3) type.
           $^{70}$Ca is the most neutron rich
           nuclei in our palette and the one for which our predictions  are less dependable because of the 
           far-off extrapolation of the neutron ESPE's. It has 
           a curious structure, more vibrational than superfluid, with its ground state wave function evenly split  (24/24/21/16)\% between the
          (0/2/4/6)p-h configurations, and a first excited 0$^+$ state at  about 500~keV of  doubly magic, N=50, Z=20, character.

 \begin{figure}[t]
\begin{center}
\includegraphics[width=\columnwidth]{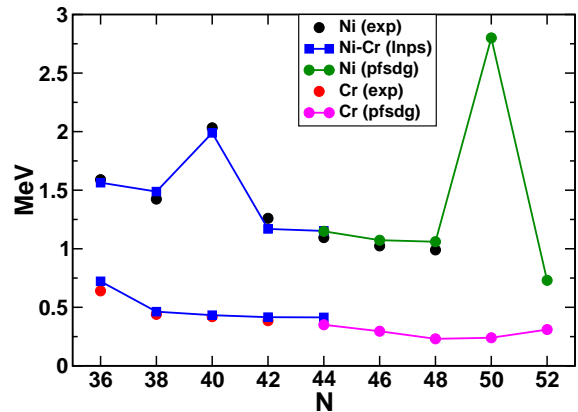}
\caption{(color online)   2$^+$  energy systematics in the
  nickel  and chromium isotopic chains.  Experimental data compared with calculations using the {\sc lnps~\cite{lnps}} and {\sc pfsdg-u}
    interactions\label{cr-nichain}}
\end{center}
\end{figure}

            Finally, we gather in Figure~\ref{cr-nichain}  the evolution of the 2$^+$ excitation energies for the nickel and chromium chains.
            The present calculations are complemented towards N=40 with the results obtained using the {\sc lnps} interaction and valence space \cite{lnps}.
            It is seen that the magic peaks at   N=40 and N=50 in the nickels disappear completely in the chromiums;   the fingerprint
            of the onset of deformation and  of the entrance in the IoI's.  The same is indeed true for the iron chain.    
            The agreement of the SM-CI description with experiment may soon extend to full chains of isotopes from the proton to the neutron drip lines, 
            for instance from 
            $^{48}$Ni and $^{44}$Cr (N=20) in the $pf$-shell
            with the {\sc kb3g} interaction, to  $^{80}$Ni and $^{76}$Cr (N=52) using {\sc pfsdg-u}.

           In conclusion, it looks as if  Nature would like to replicate the N=40 physics at N=50.  Shape
           coexistence in doubly magic   $^{78}$Ni turns out to be the portal to a new IoI at N=50, which merges with  the well established one at
           N=40  for the isotopes with Z$\le$26. With this new  addition, the archipelago of IoI's in the neutron rich shores of the nuclear chart
           counts now five members:
           \mbox{N=8, 20, 28, 40, and 50.}
           
           {\it Note. } A paper describing
            the heaviest Nickel isotopes with {\it "ab initio"} methods has been posted in arXiv~\cite{c_cluster} very recently. 
            Whereas both calculations agree in many aspects, 
            they differ in the absence in the latter of the intruder, deformed, np-nh excited band, coexisting 
            with the particle-hole states built upon the doubly magic 
            ground state of $^{78}$Ni.

{\bf Acknowledgments.} 
 This work is partly supported by MINECO (Spain) grant FPA2014-57196
and Programme  ''Centros de Excelencia Severo Ochoa''  SEV-2012-0249, and by an USIAS
Fellowship of the Universit\'e de Strasbourg.

\end{document}